\newcommand{\be}{\begin{equation}}
\newcommand{\ee}{\end{equation}}
\newcommand{\bearr}{\begin{eqnarray}}
\newcommand{\eearr}{\end{eqnarray}}

\newcommand{\eps}{\epsilon}

\newcommand{\dfrac}{\displaystyle \frac}

\newcommand{\ben}{\begin{enumerate}}
\newcommand{\een}{\end{enumerate}}
\newcommand{\bfl}{\begin{flushleft}}
\newcommand{\efl}{\end{flushleft}}
\newcommand{\ba}{\begin{array}}
\newcommand{\ea}{\end{array}}
\newcommand{\btab}{\begin{tabular}}
\newcommand{\etab}{\end{tabular}}
\newcommand{\bit}{\begin{itemize}}
\newcommand{\eit}{\end{itemize}}

\newcommand{\mpp}{M_{\pi}^2}

\newcommand{\ldue}{{\cal L}_2}
\newcommand{\lqua}{{\cal L}_4}
\newcommand{\lsei}{{\cal L}_6}
\newcommand{\mtiny}[1]{{\mbox{\tiny #1}}}
\newcommand{\MS}{\mtiny{MS}}
\newcommand{\GeV}{\mbox{GeV}}
\newcommand{\MeV}{\mbox{MeV}}

\newcommand{\per}{\;\;.}
\newcommand{\co}{\; \; ,}
\newcommand{\nn}{\nonumber \\}

\documentstyle[dina4,12pt]{article}

\begin{document}
\begin{titlepage}

\def\mytoday#1{{ } \ifcase\month \or
 January\or February\or March\or April\or May\or June\or
 July\or August\or September\or October\or November\or December\fi
 \space \number\year}

\rightline{BUTP--95/3}
\rightline{hep-ph/9502285}

\vspace*{1cm}
\begin{center} {\Large \bf Double chiral logs in the $\pi \pi$
scattering amplitude}$^\sharp$

\vspace{3cm}
{\bf G. Colangelo}

\vskip 0.5 cm
Universit\"at Bern, Sidlerstrasse 5,
CH$-$3012 Bern, Switzerland.
\vspace{2cm}

\mytoday \\
\vfill
\end{center}

\begin{abstract}
By using the Renormalization Group Equations in Chiral Perturbation
Theory, one can calculate the double chiral logs that appear at two
loops in any matrix element. We calculate them in the $\pi \pi$
scattering amplitude, where they represent the potentially largest
two loop contribution. It is shown that their correction is
reasonably small.

\noindent

\noindent

{\underline{\hspace{5cm}}}\\

\noindent
$\sharp$ Work supported in part by Schweizerischer Nationalfonds/ Bundesamt
f\"ur Bildung und Wissenschaft (BBW)/
 EEC Human Capital and Mobility Program.

\vspace{.15cm}

\end{abstract}
\end{titlepage}
\vfill \eject

\section{Introduction}

The $\pi\pi$ scattering reaction is of fundamental importance for
understanding low energy hadronic physics. Since the sixties the
most fruitful way to look at it
has been to consider the
constraints given by the symmetry properties of the strong
interactions Hamiltonian. The first predictions for the $\pi \pi$
scattering amplitude according to this method were worked out by
Weinberg in 1966 \cite{Wein66}. In recent years all the theoretical
work that goes
under the name of Current Algebra has been incorporated in a more
systematic and rigorous framework called Chiral Perturbation Theory
(CHPT) \cite{revchir}. Within this approach one assumes that the pions are
the (pseudo) Goldstone bosons of QCD, and that the singularities
generated by their exchanges dominate the Green functions at low
energy. After taking into account these singularities, CHPT works as a
systematic expansion of their residues in powers of momenta and quark
masses. Current Algebra corresponds to the leading term in this
expansion.

Gasser and Leutwyler have fully worked out the theory up to the one loop
level, and in particular  have determined all the new constants
occurring in the effective Lagrangian at this level by comparison with
experimental data \cite{GL2,GL3}. After this determination one may
calculate corrections to any Current Algebra prediction. For $\pi \pi$
scattering this was done by the same authors \cite{GLPL} who
found rather large corrections to Weinberg's predictions, even
near threshold. It is interesting to try to understand why this
happens. In fact, if we stick to threshold, we have only one expansion
parameter, the average of $u$ and $d$ quark masses, which is known to
be very small in comparison with the typical QCD scale.
After expressing this parameter through the mass and decay constant of
the pion it turns out to be
$(M_\pi/4\pi F_\pi)^2 \sim 0.01$. With such a small expansion parameter
one would expect the corrections to the leading term to be rather
small. However, the $S$-wave, $I=0$ scattering length, for example,
gets at next order a 28\% correction to the Weinberg prediction
$a_0^0=0.16$. While at first sight this may look surprising, the
reason for having such large corrections is well known \cite{LiPag}.
Beyond leading
order the unitarity property requires the appearance of nonanalytic
functions such as $\mpp \log \mpp$, which, if $\mpp$ is small, may be
significantly larger. In fact, for $a_0^0$, 90\% of the
correction is given by the term proportional to $\mpp \log
\mpp/\mu^2$, at a scale $\mu=1 \GeV$.

A natural question then
arises, whether higher orders which contribute terms like $(\mpp
\log \mpp)^n$ may still give large corrections. This is especially
interesting because the comparison with experimental data is still
not completely satisfactory, as can be seen in Table \ref{tab1}. For
$a_0^0$ in particular, large corrections would still be needed in
order to obtain agreement with the present experimental central value,
while Gasser and Leutwyler assigned a very small uncertainty (of the
order of 5\%, if one considers only the uncertainty in the value of
the low energy constants) to their one loop calculation.

On the other hand Stern and collaborators \cite{Stern1} have shown
that by allowing the quark condensate to take an unexpectedly small
value, and modifying accordingly the chiral expansion \cite{Stern2},
one could find agreement with the measurement of Rosselet et al.
\cite{Ross} of $a_0^0$ already at one loop (see Table 1 and Figure 2
in Ref. \cite{Stern1}). This makes this quantity
even more interesting, since the possibility for the quark condensate
to be numerically near to $F_\pi^3$ (and not much larger) has not
yet been clearly excluded, and this would be one of the few places
where one could find experimental evidence against or in favour of it.

Our knowledge about this quantity will most probably improve both on
the experimental and theoretical side in the near future. A full two
loop calculation is now in progress \cite{2loop}, and a new, high statistics
measurement of $K_{e4}$ decays which will be made at DA$\Phi$NE, should
sizeably reduce the present experimental error \cite{phase}.

While waiting for the full two loop calculation, which, incidentally,
is rather long and tedious, there is the
possibility to evaluate at a rather low cost the potentially more
dangerous part of the order $O(p^6)$, {\it i.e.} the double chiral
logs. The calculation can be done in a relatively simple way by using
equations that follow from general properties of the
renormalization procedure. The content of these
equations is exactly the same as that of the Renormalization Group
Equations (RGE) of a renormalizable field theory, {\it i.e.} that
(considering for example dimensional regularization) the residues of
the double pole of two loop graphs are proportional to
the residues of the single pole of one loop graphs.

The use of these equations with a nonrenormalizable Lagrangian is not
new. Weinberg derived them for the chiral effective Lagrangian in
1979 \cite{Wein79}, while several authors have used them (under the
name of pole equations) in the framework of the two dimensional
$\sigma$--model up to the four or five loop level \cite{Kazak}.

\section{Renormalization Group Equations in CHPT}

As far as we know, the only application of RGE with an effective
Lagrangian since Weinberg's original derivation \cite{Wein79}, has
been Ref. \cite{BGS}, where  these equations have been used as a check
on a full two loop calculation.

Before deriving the equations we shortly review the basic ideas and
notation of CHPT. Consider the QCD Lagrangian with two flavours in the
isospin symmetry limit $m_u=m_d=\hat{m}$, with external fields coupled
to quark bilinears:
\be
\label{LQCD}
{\cal L} = {\cal L}^0_{\mtiny{QCD}} + \bar{q} \gamma^\mu(v_\mu+a_\mu
\gamma_5)q -\bar{q}(s-i\gamma_5 p)q \co
\ee
where ${\cal L}^0_{\mtiny{QCD}}$ is the QCD Lagrangian with massless
quarks -- the masses can be absorbed in the external field $s$.

Under the assumption that the pions, because of their small mass,
dominate the low energy phenomena described by this Lagrangian, one
can construct a good representation of the generating functional of
QCD at low energy by means of an effective Lagrangian:
\be
\label{master}
\int [dq] [dA] e^{i{\cal L}} = e^{iZ[v,a,s,p]} = \int [dU] e^{i {\cal
L}_{eff}} \co
\ee
where $[dA]$ stands for the integration over the gluon fields.
The effective Lagrangian must be constructed with the only
requirements of being the most general symmetric Lagrangian containing
the pions (through the matrix $U$) and the external fields. It can be
expanded in a series of terms with increasing number of derivatives
and powers of quark masses:
\be
\label{leff}
{\cal L}_{eff} = {\cal L}_2+{\cal L}_4 +{\cal L}_6 +\ldots \per
\ee
Each term ${\cal L}_{2n}$ contains a number of monomials with a
coefficient which is not fixed by symmetry constraints. At present the
best way to determine their value is by  comparison with experimental
data. ${\cal L}_2$ contains only two arbitrary parameters, $F$ and
$B$, which are easily related to the decay constant and mass of the pion:
\be
\label{FM}
F_\pi=F(1+O(\hat{m})), \; \; \; \mpp=M^2(1+O(\hat{m})) \co
\ee
where $M^2= 2 B \hat{m}$.
At next order there are seven new free parameters. The relation
between these constants called $l_i$'s, $i=1,\ldots,7$, and measurable
quantities has been worked out by Gasser and Leutwyler in Ref. \cite{GL2}.

This expansion of the Lagrangian generates a corresponding series for
the generating functional:
\be
\label{Z}
Z[v,a,s,p]=Z_2+Z_4+Z_6+\ldots \per
\ee
As Weinberg has shown \cite{Wein79}, the contribution of loops to the
generating functional is suppressed with respect to tree diagrams:
$Z_2$ is given only by tree diagrams from $\ldue$; $Z_4$ receives
contributions from tree diagrams from $\lqua$ and one loop
diagrams with $\ldue$; $Z_6$ contains tree diagrams from $\lsei$, one
loop diagrams with one vertex from $\lqua$, and two loop diagrams
containing only $\ldue$ vertices; etc.
For more details about CHPT we refer the reader to the original
article by Gasser and Leutwyler \cite{GL2} and to recent reviews
\cite{revchir}.

Here we want to apply the RGE to the $\pi \pi$ scattering amplitude
obtained from $Z_6$. We derive the equations in a way
which is much similar to the one followed  in Ref. \cite{BGS}.
We use dimensional regularization and the minimal subtraction scheme.
Before renormalization ({\it i.e.} before including the contribution
of tree diagrams from $\lsei$), the $O(p^6)$ $\pi \pi$ scattering
amplitude looks like this:
\be
\label{a6}
A^{(6)}(s,t,u)= F^{-6}  \left[  M^{2\eps}
G(s,t,u)+ M^\eps \sum_i l_i H_i(s,t,u)
\right] \co
\ee
where $\eps=d-4$. In this expression we have simply made evident the
distinction between real two loop graphs, all incorporated into the
function $G(s,t,u)$, and the one loop graphs with one vertex from the
$\lqua$ Lagrangian, that are included into  the functions $H_i(s,t,u)$.
The dimension of the product $F^6 A^{(6)}$ is
$(\mbox{mass})^{6+2\eps}$, then, to carry out the renormalization
program we write
\be
\label{laurent}
A^{(6)}=\dfrac{\mu^{2\eps}}{F^6} \left\{ \mbox{Laurent series
of} \:\mu^{-2\eps}F^6 A^{(6)} \: \mbox{in} \: \eps \right\} \per
\ee
The curly brackets contain a function with dimension
(mass)$^6$, that will be provided by powers of external momenta and
masses. The poles in $\eps$ appearing in the Laurent series will be
cancelled by the contribution of tree graphs from $\lsei$. The
ingredients we need to analyze the structure of the two loop
amplitude are the following:
\bearr
\label{GHi}
G(s,t,u)&=&\dfrac{g_2(s,t,u)}{\eps^2} +
\dfrac{g_1(s,t,u)}{\eps}+\overline{G}(s,t,u) + O(\eps)\co \nn
H_i(s,t,u)&=&\dfrac{h_i(s,t,u)}{\eps}+\overline{H}_i(s,t,u) + +
O(\eps)  \co \nn
l_i &=& \mu^\eps \left(\dfrac{\delta_i}{\eps}+l^\MS_i + O(\eps)
\right)
\eearr
Note that $G$ and the $H_i$'s do not depend at all on $\mu$, while
the constants $l_i$'s are by definition $\mu$--independent; {\it
i.e.} we have the equations:
\be
\label{muindep}
\mu \dfrac{d}{d\mu}l_i^\MS = -\delta_i \co \; \; \; \; \; \; \;
\mu \dfrac{d}{d\mu}l_{i1}^\MS =  -l_i^\MS \per
\ee

{}From general arguments on the structure of the divergences of loop
graphs (see {\it e.g.} Ref. \cite{collins}), we know that the residues
$g_2$ and $h_i$'s are polynomials in external momenta and masses.
On the other hand, $g_1$ is in general a nonlocal function of the same
variables, due to the presence of divergent subgraphs
in the two loop graphs.
Despite the fact that CHPT is a nonrenormalizable theory, the
renormalization program can be carried out in complete analogy with
the case of a renormalizable theory. In particular, all the poles in
$\eps$ must have residues which are polynomials in the external
momenta and masses, in such a way that one can remove them by adding
counterterms to the original Lagrangian. If one writes down explicitly
the Laurent series, one immediately realizes that for the double pole
this is already obvious. In order to satisfy this property in
the case of the single pole, the nonlocal functions appearing inside
its residue must cancel. In other words, the following two equations
must be satisfied:
\bearr
g_2(s,t,u)&=& -\dfrac{1}{2}\sum_i \delta_i h_i(s,t,u) \co \nn
g_1(s,t,u)&=&-\sum_i \delta_i \overline{H}_i(s,t,u)+ \mbox{polynomial
in} \: s,t,u \; \mbox{and} \; M^2 \per
\eearr
As a consequence of the first equation
the residue of the double pole, and the coefficient of the double log
in the finite part, turn out to be proportional to the polynomial
$\sum_i \delta_i h_i(s,t,u)$.

With a relatively simple calculation we are then able to get
interesting information about the two loops. To have the complete
$O(p^6)$ result one still has to
calculate part of the coefficient of the single log, and the finite
contribution.

\section{Double logs in the $\pi \pi$ scattering amplitude}

We have calculated the polynomials $h_i$ partly
using Feynman diagrams and partly with the Heat
Kernel technique. In some cases we have performed the
calculation with both methods in order to have a cross check. More
details about the calculation can be found in Ref. \cite{tesi}. Here
we simply give the results:
\bearr
\label{his}
h_1(s,t,u)&=&\frac{1}{16\pi^2 }\left[-\frac{21}{2}s^3
-\frac{5}{6} s(t-u)^2+\frac{200}{3} M^2 s^2
 -\frac{4}{3} M^2 (t-u)^2 \right. \nn
&& \; \; \; \; \; \; \; \; \;
\left.-164 M^4 s +\frac{376}{3} M^6
\right] \co \nn
h_2(s,t,u)&=&\frac{1}{16\pi^2 }\left[-\frac{107}{12}s^3
-\frac{25}{12} s(t-u)^2+ 44 M^2 s^2 -\frac{4}{3}M^2 (t-u)^2 \right. \nn
&& \; \; \; \; \; \; \; \; \;
\left. -\frac{296}{3}M^4 s +\frac{208}{3}M^6
\right] \co \nn
h_3(s,t,u)&=&\frac{1}{16\pi^2}\left[-4 M^4 s -2 M^6\right] \per
\eearr
All the other $h_i$'s are zero.
Before using these expressions to get numerical results, we need to
shuffle part of the two loop amplitude $A^{(6)}$ into the lower order
amplitudes in order to express everything in terms of the physical
pion mass and the physical decay constant. Since we are calculating
only the divergent part of one loop graphs containing one $l_i$ in
$A^{(6)}$, we need to calculate the same contribution to $\mpp$ and
$F_\pi$:
\bearr
\label{MF}
M_\pi^2 &=&M^2\left\{1+\frac{M^2}{
F^2}\left[2l_3+ \frac{1}{2 i} \frac{\Delta(0)}{M^2} \right] \right.\nn
&& \; \; \; \; \; \; \; \; \; \;
\left.
+ \frac{M^4}{
F^4}\left[(-14l_1-8l_2-3l_3) \frac{1}{i}\frac{\Delta(0)}{M^2} +\ldots
\right] +O(M^6)\right\} \co \nn
F^2_\pi&=& F^2\left\{1+\frac{M^2}{
F^2}\left[2l_4-\frac{2}{i} \frac{\Delta(0)}{M^2} \right] \right.\nn
&& \; \; \; \; \; \; \; \; \; \;
\left.
+ \frac{M^4}{
F^4}\left[(14 l_1+8l_2-4l_3-3l_4) \frac{1}{i}\frac{\Delta(0)}{M^2}+\ldots
\right] +O(M^6)\right\} ,
\eearr
where $\Delta(z)$ is the Feynman propagator of a scalar field with
mass $M$: $\Delta(0)= 2i M^\eps/(16\pi^2 \eps) + O(1)$.

In order to evaluate the size of these corrections to the $\pi \pi$
scattering amplitude, we calculate their
contribution to threshold parameters, following the conventions of
Ref. \cite{GL2}\footnote{The ellipses between parentheses stand
for the one loop contributions that we do not display here
in order not to put too many formulae in this note. We refer the
reader again to Ref. \cite{GL2} (note that the corrections there are
usually expressed in terms of $M$ and $F$ and {\it not} with the
physical mass and decay constant, as we do here).}:
\bearr
\label{scattlen2}
a^0_0 &=& \frac{7\mpp}{32 \pi F_\pi^2} \left\{1+ (\ldots)+
\frac{M_\pi^4}{F_\pi^4} \left[-\frac{58}{7}k_1
-\frac{96}{7}k_2-5 k_3-\frac{11}{2} k_4 +\ldots\right]
+O(M_\pi^6) \right\} \co \nn
&& \nn
b^0_0 &=& \frac{1}{4\pi F_\pi^2}\left\{ 1+
(\ldots) +\frac{M_\pi^4}{F_\pi^4}\left[ -\frac{86}{3}k_1-\frac{73}{2}
k_2 -\frac{5}{2}k_3 -\frac{29}{3}k_4+\ldots \right] + O(M_\pi^6)
\right\} \co \nn
&&\nn
a^2_0&=&-\frac{\mpp}{16\pi F_\pi^2} \left\{ 1+
(\ldots)+\frac{M_\pi^4}{F_\pi^4} \left[2
k_1+6 k_2 + k_3 + \frac{1}{2}k_4+\ldots \right]
+O(M_\pi^6) \right\} \co \nn
&& \nn
b^2_0&=&-\frac{1}{8\pi F_\pi^2} \left\{1+(\ldots)+
\frac{M_\pi^4}{F_\pi^4} \left[
-\frac{20}{3}k_1+\frac{3}{2}k_2-\frac{5}{2}k_3 +\frac{7}{3}k_4 +\ldots
\right] +O(M_\pi^6) \right\} \co \nn
&& \nn
a^1_1 &=&\frac{1}{24\pi F_\pi^2} \left\{1+
(\ldots)+\frac{M_\pi^4}{F_\pi^4}
\left[ -12 k_1 -\frac{119}{6}k_2-\frac{5}{2}k_3- 3k_4
  +\ldots \right]
+O(M_\pi^6) \right\} \co \nn
b^1_1&=&\frac{1}{ F_\pi^4}(\ldots)
+\frac{\mpp}{18\pi F_\pi^6}\left[-23
k_1-\frac{109}{4}k_2+\ldots \right]+O(M_\pi^4)\co \nn
&& \nn
a^0_2&=&\frac{1}{F_\pi^4} (\ldots)+\frac{\mpp}{10\pi
F_\pi^6}\left[\frac{11}{18} k_1 -\frac{8}{3}k_2 - \frac{5}{9}k_4
+\ldots \right] +O(M_\pi^4)\co \nn
&& \nn
a^2_2 &=&\frac{1}{F_\pi^4}(\ldots)+\frac{\mpp}{10\pi
F_\pi^6}\left[\frac{16}{9}k_1 + \frac{3}{4}k_2- \frac{2}{9}k_4 +\ldots
\right] +O(M_\pi^4) \co
\eearr
where
\be
k_i=\left(4 l_i^r -\delta_i \log\frac{\mpp}{\mu^2}\right)
\frac{1}{16\pi^2} \log \frac{\mpp}{\mu^2}\co \; \; \; \; \; \;
i=1,\ldots,4 \per
\ee
Note that with this definition of the $k_i$'s we are also taking into
account in Eqs. (\ref{scattlen2}) the part of the single log
which is proportional to the renormalized constants. From our previous
considerations one can see that this contribution is unambiguously
determined by the polynomials $h_i$.
Note also that the renormalized constants we are using here are
defined in the renormalization scheme adopted by Gasser and Leutwyler
in Ref. \cite{GL2}, and not in the MS scheme that we used in the
previous section to simplify the notation.

Before analyzing the part of the two loop contribution evaluated here
it is interesting to investigate in some more details the structure of the
$O(p^4)$ correction. In the third column of Table \ref{tab1}, we show how
much of that correction is due to the chiral log evaluated at
a scale $\mu=1 \GeV$. It is clear that for the $S$-wave parameters the
chiral log is responsible for almost all the correction. On the other
hand, for the $P$-wave, the chiral log arises only through $F_\pi$
renormalization -- had one used the pion decay constant in the chiral
limit $F$, there would not have been any log (and in fact for $b_1^1$,
which does not contain any tree level contribution, so that the use of
$F$ or $F_\pi$ is equivalent, there is no chiral log at this order).
Moreover, in this case we expect the
$\rho$ to give an important contribution, and it is well known that the
resonances manifest themselves through the $l_i^r$'s. Hence the fact
that the chiral logs are not the main part of the one loop correction
does not come out as a surprise.
Finally we have the two $D$-wave scattering lengths, for which the
order $O(p^6)$ is the next to leading correction. In this case we feel
that it is more hazardous to estimate it only through the double
chiral logs; also because, a posteriori, the part of the single chiral
logs that we get turns out to be as important as the double logs.
Moreover the $I=2$ scattering length at leading order is unnaturally
small because of a strong cancellation between the chiral log and the
finite part.

{}From all these observations we conclude that for $P$ and $D$ waves the
assumption that the double chiral logs could be a good estimate of the
full two loop correction is poorly justified.

The numerical results are displayed in Table \ref{tab1}, together with
the one loop predictions and experimental data available. Since the
numbers we produce are $\mu$-dependent we have calculated them
for two reasonable values of $\mu$.  Moreover, in Table \ref{tab1} we
show both the double logs and the complete $k_i$'s contributions.

We observe that:
\ben
\item
For the $S$-wave parameters this $O(p^6)$ correction is, in the
extreme case,  of the order of 10\%. To judge whether the size of this
correction is reasonable or not, we may compare the coefficients of
the single and double logs in the $O(p^4)$ and $O(p^6)$ corrections,
respectively:
\bearr
\label{LE}
a^0_0 &=& \frac{7\mpp}{32 \pi F_\pi^2} \left\{1 - \frac{9}{2}L +
\frac{857}{42} L^2 + \ldots \right\} \co \nn
&& \nn
b^0_0 &=& \frac{1}{4\pi F_\pi^2}\left\{ 1 -\frac{26}{3} L +
\frac{1871}{36} L^2 +\ldots
\right\} \co \nn
&&\nn
a^2_0&=&-\frac{\mpp}{16\pi F_\pi^2} \left\{ 1+ \frac{3}{2} L -
\frac{31}{6}L^2 + \ldots \right\} \co \nn
&& \nn
b^2_0&=&-\frac{1}{8\pi F_\pi^2} \left\{1+\frac{10}{3}L - \frac{83}{18}
L^2 + \ldots \right\} \co
\eearr
where
$L\equiv (\mpp/16 \pi^2 F_\pi^2) \log \mpp /\mu^2$.
In first place one can see that the absolute value of the $L^2$
coefficient is numerically quite similar to the square of the $L$
coefficient, in all cases but $b_0^2$. [Incidentally, had $b_0^2$
respected this rule of thumb for the absolute value, with the sign it
has, the distance from the present experimental central value would
have increased sizeably.]
The second interesting point is that they show a very regular sign
pattern. In order to have some more insight into it, it is useful to
disentangle two effects of comparable importance: the correction to
the amplitude $A(s,t,u)$ expressed in terms of $F$, and the effect of
the renormalization of $F$ into $F_\pi$. This is done in Table
\ref{tab3}, where one can see that whilst the two effects add up for
the $I=0$ parameters, they partially cancel in the case of the $I=2$
parameters. In particular, in the latter case one can trace back the
minus sign of the coefficient of $L^2$ to the interference of the two
one loop corrections.
\item
As far as the $P$- and $D$-wave parameters are concerned, we have
displayed here the double chiral logs contribution, essentially for
the sake of completeness. As we stressed already, we see no special
reasons why the numbers we find here should be representative of the
full two loop correction.
\item
As we mentioned in the introduction, it is hoped that there will be an
improvement in the experimental knowledge of $a_0^0$ in the near
future, arising from a new measurement of $K_{e4}$ decays. On the
theoretical side the full two loop calculation, which is now in
progress \cite{2loop}, will improve the one loop result of Gasser and
Leutwyler \cite{GL2,GLPL}. Our results show that the contribution from
the two loops is expected to be reasonably small, and confirm that it looks
very unlikely that CHPT could at any order reproduce the present
experimental central value.
\een

\vspace{1.5cm}

\noindent
{\bf{Acknowledgements}}\\
It is a pleasure to thank J\"urg Gasser for his help and advice
throughout this work and a careful reading of the manuscript, and
Heiri Leutwyler for interesting discussions. I wish to thank also
Hans Bijnens and Gerhard Ecker for the fruitful collaboration we are
having in the enterprise of calculating the two loops.

\vspace{1.5cm}

\newcommand{\PL}[3]{{Phys. Lett.}        {#1} {(19#2)} {#3}}
\newcommand{\PRL}[3]{{Phys. Rev. Lett.} {#1} {(19#2)} {#3}}
\newcommand{\PR}[3]{{Phys. Rev.}        {#1} {(19#2)} {#3}}
\newcommand{\NP}[3]{{Nucl. Phys.}        {#1} {(19#2)} {#3}}

\clearpage
\newpage

\begin{table}[htb]
\caption{Numerical values of the threshold parameters according to  Eqs.
(\protect\ref{scattlen2}), for
$\mu=0.5, \; 1\protect\GeV$. In the column containing the one loop
results, the numbers between parentheses show how much of the one loop
contribution is due to the chiral log evaluated at the scale $\mu=1
\GeV$. In the calculation we have used
$M_\pi=139.6 \MeV$ and $F_\pi=93.2 \MeV$. To calculate the $k_i$'s we
have used the $l_i^r$'s given in Table \protect\ref{tab2}. Scattering lengths
$a^I_l$, and effective ranges $b_l^I$ are given in the appropriate
powers of $M_{\pi^+}$. The value of $\mu$ is expressed in GeV.}
\label{tab1}
\begin{center}
\vspace{.5cm}
\begin{tabular}{|c|cr|cc|cc|c|}
\hline
 & \multicolumn{2}{c|}{1 loop $\; \; (\log)$} & \multicolumn{2}{c|}{1
loop +$\log^2 \mpp$}&\multicolumn{2}{c|}
{ 1 loop+$k_i$'s} & experiment   \\
&        & $\mu=1$&$\mu =0.5$ & $\mu=1$ & $\mu =0.5$  &
$\mu=1$  &\cite{nagels}\\
\hline
$ a_0^0$      &$0.201$ & (88\%)  &$0.205$  &$0.211$ &$0.211$  &$0.213$
&$0.26\pm0.05$\\

$b_0^0$       &$0.248$ & (125\%) &$0.260$  &$0.277$ &$0.275$  &$0.279$
&$0.25\pm0.03$\\

$-$10 $a_0^2$ &$0.418$ & (133\%) &$0.415$  &$0.411$ &$0.409$  &$0.407$
&$0.28\pm0.12$\\

$-$10 $b_0^2$ &$0.726$ & (100\%) &$0.721$  &$0.713$ &$0.701$  &$0.691$
&$0.82\pm0.08$\\

$10 a_1^1$    &$0.370$ & (46\%)  &$0.379$  &$0.390$ &$0.391$  &$0.395$
&$0.38\pm0.02$\\
\hline
$10^2 b_1^1$  &$0.485$ & (0\%)   &$0.619$  &$0.805$ &$0.749$  &$0.780$
&\\

$10^2 a_2^0$  &$0.181$ & (122\%) &$0.206$  &$0.241$ &$0.273$  &$0.302$
&$0.17\pm0.03$ \\

$10^3 a_2^2$  &$0.205$ & (432\%) &$0.145$  &$0.061$ &$0.267$  &$0.335$
&$0.13\pm0.3$\\
\hline
\end{tabular}
\end{center}
\end{table}

\begin{table}[hb]
\caption{Numerical values of the  coupling constants $l_i^r$'s and the
$k_i$'s used in Table \protect\ref{tab1}. The values
of $l_3^r$ and $l_4^r$ are taken from Ref. \protect\cite{GL2}, while
for the other two we have used the more recent ones obtained in Ref.
\protect\cite{BCG}. The value of $\mu$ is expressed in GeV.}
\label{tab2}
\begin{center}
\vspace{.5cm}
\begin{tabular}{|c|rr|rr|}
\hline
i  & \multicolumn{2}{c|}{$10^{3}l_i^r$}&\multicolumn{2}{c|}{$10^{3} k_i$}  \\
       & $\mu =0.5$  & $\mu=1$ & $\mu =0.5$  &
$\mu=1$ \\
\hline
1 & -4.5 &-5.9 & 0.20 & 0.39 \\
2 &  7.5 & 4.6 &-0.65 &-0.87 \\
3 & -0.5 & 1.6 & 0.17 & 0.15 \\
4 & 11.1 & 2.3 &-1.24 &-1.47 \\
\hline
\end{tabular}
\end{center}
\end{table}

\begin{table}[hb]
\caption{Breakdown of the coefficient of $L$ and $L^2$ into
contributions arising from two different effects: the corrections to
the amplitude $A(s,t,u)$, and the renormalization of $F$ into $F_\pi$.
For the $L^2$ coefficient we have also separated out the contribution
due to the interference between the two one loop effects. [The small
effect due to mass renormalization is included into the columns tagged
by $A(s,t,u)$.]}
\label{tab3}
\begin{center}
\vspace{.5cm}
\begin{tabular}{|c|c|c|c||c|c|c|c|}
\hline
  &\multicolumn{3}{c||}{$L$-coefficient}& \multicolumn{4}{c|}{
$L^2$-coefficient}  \\
  &\multicolumn{3}{c||}{}& \multicolumn{4}{c|}{}\\
\hline
 & $A(s,t,u)$& $F_\pi$ & total & $A(s,t,u)$&interference & $F_\pi$&
total \\
&&&&&&&\\
\hline
&&&&&&&\\
$a_0^0$ & $-\dfrac{5}{2}$ & $-2$ & $-\dfrac{9}{2}$ & $\dfrac{145}{21}$ & 10
& $\dfrac{7}{2}$ & $\dfrac{857}{42}$ \\
&&&&&&&\\
$b_0^0$ & $-\dfrac{20}{3}$ & $-2$ & $-\dfrac{26}{3}$ & $\dfrac{785}{36}$ &
$\dfrac{80}{3}$ & $\dfrac{7}{2}$ & $\dfrac{1871}{36}$ \\
&&&&&&&\\
$a_0^2$ & $\dfrac{7}{2}$ & $-2$ & $\dfrac{3}{2}$ & $\dfrac{16}{3}$ & $-14$
& $\dfrac{7}{2}$ & $-\dfrac{31}{6}$ \\
&&&&&&&\\
$b_0^2$ & $\dfrac{16}{3}$ & $-2$ & $\dfrac{10}{3}$ & $\dfrac{476}{36}$ &
$-\dfrac{64}{3}$ & $\dfrac{7}{2}$ & $-\dfrac{83}{18}$ \\
&&&&&&&\\
\hline
\end{tabular}
\end{center}
\end{table}

\end{document}